# Reinforcement of interfacial superconductivity in a $Bi_2Te_3$/$Fe_{1+y}Te$ heterostructure under hydrostatic pressure

Junying Shen[a], Claire Heuckeroth[b], Yuhang Deng[b], Qinglin He[a*], Hong Chao Liu[a†], Jing Liang[a], Jiannong Wang[a], Iam Keong Sou[a], James S. Schilling[b] and Rolf Lortz[a*‡]

[a] Department of Physics, The Hong Kong University of Science and Technology, Clear Water Bay, Kowloon, Hong Kong

[b] Department of Physics, Washington University, CB 1105, One Brookings Dr., St. Louis, MO 63130, USA

## Abstract

We investigate the hydrostatic pressure dependence of interfacial superconductivity occurring at the atomically sharp interface between two non-superconducting materials: the topological insulator (TI) $Bi_2Te_3$ and the parent compound $Fe_{1+y}Te$ of the chalcogenide iron based superconductors. Under pressure, a significant increase in the superconducting transition temperature $T_c$ is observed. We trace the pressure dependence of a superconducting twin gap structure by Andreev reflection point contact spectroscopy (PCARS), which shows that a large superconducting gap associated with the interfacial superconductivity increases along with $T_c$. A second smaller gap, which is attributed to proximity-induced superconductivity in the TI layer, increases first, but then reaches a maximum and appears to be gradually suppressed at higher pressure. We interpret our data in the context of a pressure-induced doping effect of the interface, in which charge is transferred from the TI layer to the interface and the interfacial superconductivity is enhanced. This demonstrates the important role of the TI in the interfacial superconductivity mechanism.

[*] Present address: Department of Electrical Engineering, University of California, Los Angeles, California 90095, USA.

[†] Present address: School of Physics and Astronomy, University of Birmingham, Birmingham B15 2TT, UK.

[‡] Corresponding author. Email: lortz@ust.hk

# 1. Introduction

Topological insulators (TI) have been a hot research topic since predicted and discovered in 2007 [1,2]. The novel three-dimensional TIs, e.g. $Bi_2Te_3$ and $Bi_2Se_3$, own strong topological surface states with helical Dirac fermions, protected by time-reversal symmetry. Meanwhile, interfacial superconductivity also attracts intense interest from both theorists and experimentalists in various materials [3-5]. By combining a TI with an s-wave superconductor, the proximity effect at the surface states of TI has been studied, suggesting the existence of a topological superconductor, which may hold Majorana fermionic states [6]. This idea has triggered a search for Majorana fermions [7,8], which are of high interest for future applications in quantum computation [9].

We have recently reported a novel $Bi_2Te_3/Fe_{1+y}Te$ heterostructure, which exhibits interfacial superconductivity at an atomically-flat van-der Waals-bonded boundary between the non-superconducting parent compound $Fe_{1+y}Te$ of the '11' iron-based superconductor family and the non-superconducting topological insulator $Bi_2Te_3$ [10]. In this heterostructure, the TI surface state is combined with the electronic complexity of the iron-based chalcogenide materials resulting in a complex interfacial superconducting state. The interface exhibits all characteristic fingerprints of two-dimensional (2D) superconductivity with a Berezinski-Kosterlitz-Thouless (BKT) transition below the resistive onset critical temperature ($T_c^{onset}$). In addition, a square-root or linear temperature dependence of the parallel and perpendicular upper critical fields, respectively, was observed, which follow the 2D Ginzburg-Lanudau theory for a superconducting thickness of 7 nm. While a pure $Fe_{1+y}Te$ film does not show any sign of superconductivity, with a $Bi_2Te_3$ film coated on the $Fe_{1+y}Te$ the thin interfacial layer of the heterostructure becomes superconducting starting from one quintuple layer (QL) thickness of the TI. The $T_c^{onset}$ increases with further $Bi_2Te_3$ thickness until it saturates at about 12 K for thicknesses exceeding 5 QLs. Since 5 QLs has been reported to represent the lower thickness limit of $Bi_2Te_3$ in order to fully develop a topological surface state [11,12], the dependence of $T_c^{onset}$ on the number of QLs suggests that the TI surface states likely play a crucial role for the emergence of the interfacial superconductivity. In addition, it has been reported that at the interface between $Bi_2Te_3$ and $Fe_{1+y}Te$ the magnetic ordering of the latter compound is altered so that the Fe magnetic moments order out of plane and thus perpendicular to the interface [13].

However, the superconducting mechanism remains unknown. The complexity of the superconducting gap structure, as observed in our previous point contact spectra [14] with multiple superconducting gaps, a pronounced pseudogap up to 30 K above $T_c^{onset}$, and a robust zero bias conductance peak, suggest a complex and peculiar superconducting pairing mechanism. In addition, it is unclear whether the highest critical temperature observed for more than 5 QLs thickness of $Bi_2Te_3$ can be further increased by varying the charge carrier concentration, for example by ionic substitution or application of external pressure.

Point contact Andreev-reflection spectroscopy (PCARS), as an energy-resolved technique, is widely used to investigate superconductors ever since it was first introduced by Yanson [15]. With such a technique, not only the amplitude and symmetry of the order parameter, but also the nature of the pairing boson, or even a hint of the shape of the fermion surface may be obtained [16]. In a previous work [14], we focused on a heterostructure with a TI thickness of 9 QLs (9 nm) and a resistive $T_c^{onset}$ of 12 K at ambient pressure, using a directional point-contact technique, which revealed two superconducting gaps together with a large pseudogap persistent up to 40 K. By probing from the top surface of the TI with a scanning probe tip and from the interface edge with a nano-contact method, the PCARS clearly indicated an isotropic smaller gap (~6 meV) associated with the TI and an anisotropic larger gap (~12 meV) associated with a thin layer of $Fe_{1+y}Te$ in the vicinity to the interface.

High pressure has been known to be a powerful technique to induce superconductivity [17] and to modify the lattice structure and the charge carrier concentration [18,19]. The latter allows experiments in which an unconventional (e.g. iron-based superconductor) is tuned from the underdoped regime towards the overdoped side of the phase diagram as a function of charge carrier concentration on one single stoichiometric sample, without introducing any crystalline disorder. In this article, we report data of PCARS on the $Bi_2Te_3/Fe_{1+y}Te$ (9 QLs) heterostructure in combination with electrical resistivity and under the influence of hydrostatic (He-gas) pressure. This allows us to directly study the pressure evolution of the superconducting gaps, a method, which is rarely performed under pressure. A clear enhancement of both $T_c^{onset}$ and the temperature $T_o$ where zero resistivity is established is observed in the resistance as the pressures increases up to 5.9 Kbar. In addition, a sharp resistivity peak that occurs just above $T_c^{onset}$ observed at ambient pressure is suppressed. PCARS were obtained under different pressures from ambient conditions up to 6.7 Kbar. By fitting our data to a modified Blonder-Tinkham-Klapwijk (BTK) model [16,20], we are able to trace the pressure dependence of the two superconducting gaps. The pressure data is in agreement with a pressure-induced increase of the interfacial charge carrier concentration, and suggests that our interface at ambient pressure is still in the underdoped regime and has the potential for an even higher critical temperature.

## 2. Experimental Details

The heterostructure studied in this work was synthesized in a VG-V80H MBE system. A featured thickness of 9 quintuple layers (9 nm) of $Bi_2Te_3$ was capped onto the $Fe_{1+y}Te$ layer (140 nm). The latter was grown on top of a ZnSe buffer layer (50 nm) deposited on the GaAs (001) semi-insulating substrates. Further details on the sample growth, morphology and characterization can be found in Ref. 10.

The electrical resistance and point contact measurements were conducted in a helium cryogenic Janis Supervaritemp bath cryostat using the standard four-probe method. For the resistance measurements an AC excitation current of 0.1 µA at 77 Hz was used. The sample voltage was amplified by a Stanford Research SR554 transformer preamplifier and fed into an SR839 DSP lock-in amplifier. Each sample was cut into the form of a long strip and silver paint was used to fabricate electrodes on the top surface of $Bi_2Te_3$. Various attempts to establish a nano-size contact at the edge of the sample, as we reported at ambient pressure previously [10], failed because the contact was unstable upon application of pressure. In order to realize more reliable point-contacts under pressure, we adopted the simpler but well-established 'soft' point-contact technique, in which the contact is formed by a drop of silver paint (with a diameter of 100 um) on the clean top surface of the heterostructure [14]. Even though the typical diameter of such a silver paint drop is much larger than the coherence length (~5 nm [10]), only parallel nanometric channels of size less than 10 um are formed by the individual Ag grains in the paint [21]. Since the silver paint is known to quickly diffuse into $Bi_2Te_3$ [22,23], our electrodes should contact the entire TI layer and also a part of the $Fe_{1+y}Te$ layer below. Thus, we are effectively probing both layers simultaneously. Alternatively to the 'Needle-anvil' method, the 'soft' point-contact technique eliminates the lattice distortion induced by pressure from the contact, meanwhile, it ensures better mechanical and thermal stability, avoiding thermal drift of the tip, especially in the high pressure environment. A drawback is that the barrier height $Z$ of the point contact (the contact resistance) cannot be controlled during the experiment. This may lead to a random variation of the spectra between the Andreev-reflection limit and the tunneling limit at different pressures. For technical reasons the measurements have been conducted at 4.2 K. The spectra thus appear broader than the ambient pressure data in Ref. 14, where lower temperatures down to 300 mK were available. However, the BTK model can easily account for both the variation of $Z$ and the broadening, and allowed us to extract the pressure dependence of the two superconducting gaps with sufficient precision.

To generate the pressure up to 7 kbar a He-gas compressor (Harwood Engineering) was connected to a CuBe helium gas pressure cell (Unipress, Warsaw) via a long capillary. With the sample sealed inside, the CuBe pressure cell was immersed in the liquid helium bath cryostat. The pressure values displayed in this article were measured using a digital manganin gauge at ambient temperature connected to the pressure vessel by a thin CuBe capillary tube. At pressures below the freezing point of helium a correction was made using known helium isochores to obtain the slightly lower values of the pressure at 4 K. In this experiment the pressure was only changed at ambient temperature. Further details of the high-pressure experiment can be found in Ref. 24.

## 3. Results

Fig. 1. shows the temperature-dependent resistance of the $Bi_2Te_3$/$Fe_{1+y}Te$ heterostructure (9 QLs) on a logarithmic temperature scale. Data at ambient pressure have been reported previously in Ref. [10,14]. Upon cooling from 295 K, the resistance is dominated by the bulk $Fe_{1+y}Te$, which effectively shunts the much thinner interface layer. It displays the characteristic insulator-to-metal transition with a symbolic bump located at 76 K, characteristic for $Fe_{1+y}Te$ with a rather high excess Fe content [25]. This transition is well-known to originate from the antiferromagnetic double-stripe spin density wave ordering in bulk $Fe_{1+y}Te$ [26], and the corresponding characteristic temperature is noted as $T_{SDW}$. Note that $Fe_{1+y}Te$ does not exist in a stoichiometric form (FeTe), but always contains excess Fe in the form of interstitial Fe. In our heterostructure the value of $y$ is $0.15 \pm 0.02$, as indicated by scanning transmission electron microscopy [10].

Upon further cooling the ambient pressure resistance goes through a minimum at 24 K, followed by a sudden increase with a sharp maximum at around 12 K, denoted as the onset $T_c$ ($T_c^{onset}$), then a superconducting transition is clearly seen as the resistance drops gradually to zero at $T_o$. This dramatic peak at $T_c^{onset}$ is also typical for $Fe_{1+y}Te$ with high excess Fe [25], and likely originates from scattering on the interstitial Fe magnetic moments in the bulk $Fe_{1+y}Te$ layer. The superconducting transition has been demonstrated to be restricted to a 7-nm-thin planar region in the vicinity of the interface [10]. The resistance of the heterostructure is actually represented by three resistances is parallel: the thick bulk $Fe_{1+y}Te$ layer, the interface layer and the $Bi_2Te_3$ layer. Above the superconducting transition onset the interface is normal conducting and the total resistance will be dominated by the much thicker bulk $Fe_{1+y}Te$ layer. Below the superconducting onset the superconducting interfacial layer will then shunt the $Fe_{1+y}Te$ and $Bi_2Te_3$ layers.

Upon application of pressure up to 5.9 kbar it is obvious that pressure decreases the overall $Fe_{1+y}Te$ normal state resistance and suppresses the steep resistance increase just above $T_c$. In addition, the superconducting transition of the interface is clearly enhanced and both, $T_c^{onset}$ and the zero resistance temperature $T_o$ are significantly increased. This is illustrated in the inset of Fig. 1, where both $T_c^{onset}$ and $T_o$ are plotted against the applied pressure. Note that pure $Fe_{1+y}Te$ is not superconducting under hydrostatic pressure conditions [27], therefore the superconductivity is attributed at all pressures to the interface layer. It has been demonstrated previously that the finite transition width of the superconducting transition at ambient pressure is a consequence of the 2D nature of superconductivity, and well described by a BKT transition with additional finite size effects that can significantly alter the transition width [10]. In addition, the transition width increases drastically when the heterostructures are stored under protective nitrogen gas for several months, which has been found to be intrinsic and not due to a deterioration in the sample quality [28]. This effect is likely a consequence of the ordering processes of interstitial magnetic iron ions, and we will provide further evidence in this paper. The latter is the case for the investigated heterostructure, which enters a zero-resistance state only below $T_o = 4$ K where the superconductor becomes globally phase coherent. However, its normal state properties and the transition onset are identical to those previously reported [10,14].

The peak above $T_c^{onset}$ is much lowered and almost flattened with a pressure of 3.9 kbar, then it disappears at 5.9 kbar together with a sudden displacement of $T_c^{onset}$ up to 20 K. Meanwhile, $T_o$ increases to 11.6 K at a pressure of 2.13 kbar and reaches $T_o = 12.3$ K at 3.9 kbar.

To investigate the pressure development of the superconducting gaps, we investigated the PCARS performed on $Bi_2Te_3/Fe_{1+y}Te$ under He-gas pressures at 4.2 K after the pressure was applied at room temperature. Seven groups of data were taken at different pressures from ambient pressure up to 6.7 kbar. We show all data at different pressures (normalized by a very smooth, featureless concave polynomial background fitted in the high bias voltage range for clarity) in Fig. 2 a-g (open circles). Since the point-contact technique is a combination of tunneling effects and Andreev reflections, in which the former dominate in large-resistance contacts and the latter in smaller ones, the point-contact spectra strongly depend on the contact resistance, which determines the barrier height $Z$. We have spectra with a dip-like structure (Fig. 2 d) typical of the tunneling limit, but also peak structures at other pressures (Fig. 2 a, b, f and g) indicating the Andreev reflection case, as well as spectra in the intermediate regime (Fig. 2 c and e). This is a consequence of the pressure dependence of the point contact, and subtle changes in the conditions can lead to large changes in the arrangement of the nanometric channels in the sliver paint spot. The BTK model describes point contacts in both limits and in between, taking into account $Z$ as a fitting parameter.

To fit the data, we used a 2-gap s-wave BTK model and the fits are included in Fig. 2 (solid lines). We have also attempted to fit with a 2-gap d-wave BTK model but no significant difference has been found due to the rather high temperature and the geometry of the contact, in which tunneling occurs in the direction perpendicular to the layer structure. Therefore, only the results of the s-wave fits are provided in Fig. 2 unless otherwise specified. The BTK model fits very well with the data. An exception is the 2.2 kbar data (Fig. 2b), in which the very sharp zero-bias conductance peak can not be reproduced by an s-wave model. It agrees with the sharp peaks observed in the Andreev limit and the zero bias conductance peak in the tunneling limit in the data reported at ambient pressure [14]. It may suggest a nodal order parameter that has been measured with a contact in which the current was injected with a small in-plane component along the node direction. We have added a fit where the smaller gap is replaced by a d-wave gap in the tunneling limit along the node direction of identical gap amplitude, and the fit describes the data including the zero bias peak very well. The possible nodal gap structure may be the fingerprint of a complex topological nature of the proximity-induced superconductivity in the $Bi_2Te_3$ layer. While this stimulates further work to clarify the pairing symmetry, we will first focus on the pressure evolution of the superconducting gaps in this article.

At ambient pressure the data is clearly in the Andreev limit, and a broad peak appears with two small dip-like features at ±5 mV. Although the spectrum is broader than reported for lower temperatures in Ref. 14, we can clearly distinguish the two superconducting gaps. The larger one is forming the broad background peak and the smaller one being framed by the dip-like features. The fitting results provide gap values of 2 meV ($\Delta_1$) and 8 meV ($\Delta_2$) in excellent agreement with our earlier work at a similar temperature [14].

At 2.2 kbar pressure, the contact remains in the Andreev limit and the spectrum appears rather similar to the ambient pressure data. The broad peak associated with the large gap remains virtually unchanged, while the dip-like features surrounding the smaller gap transform into a gentle upturn in the direction of the sharp zero-bias conductance peak. The smaller gap $\Delta_1$ increases to 2.6 meV and $\Delta_2$ to 10 meV.

When the pressure is increased to 3.2, 4.0 and 4.9 kbar, the spectra obtain some tunneling characteristics. At 3.2 kbar the large gap is seen as positive feature, which is mainly generated by Andreev reflections. The smaller gap changes into the tunneling limit and appears as a dip in the middle. At 4.0 kbar, the contact is completely in the tunneling limit and a clear tunneling gap with shoulder-like structures is observed, indicating the presence of the two gaps of 4.2 meV ($\Delta_1$) and 13.2 meV ($\Delta_2$). The data recorded at 4.9 kbar resembles the data at 3.2 kbar, with the large gap having a positive Andreev signature and the small gap appearing as a dip structure. There are sharper shoulder-like structures that indicate an intermediate case between the Andreev regime and the tunneling limit. At 5.3 kbar, a broad peak characteristic for the Andreev regime appears again, with the small gap in the form of a very shallow dip visible in the low-bias voltage regime on top of the broad peak. Finally, a broad Andreev peak can be seen at 6.7 kbar. For the latter, it is difficult to distinguish the two gaps with the eye. However, the BTK model is still able to extract the two individual values of the gaps, which together cause the characteristic pyramid shape of the spectrum.

## 4. Discussion

The pressure development of each gap obtained from the BTK fits is shown in Fig. 2h. $\Delta_2$ increases continuously as a function of the pressure from 8 meV at ambient pressure up to 15 meV at 6.7 kbar. The smaller gap $\Delta_1$ initially increases from 2 meV at ambient pressure to a value of 4.3 meV at 4.9 kbar, where it appears to saturate and then slightly decreases. The continuous increase of $\Delta_2$ (associated with the superconducting $Fe_{1+y}Te$ layer at the interface [10]) agrees with the observed increase of $T_c$ in the resistivity data. The initial increase in the smaller gap $\Delta_1$ (associated with proximity induced superconductivity in the $Bi_2Te_3$ layer [10]) appears to follow this trend, presumably due to a reinforcement of the proximity effect when $\Delta_2$ grows, but this trend does not continue in the higher pressure regime above 5 kbar, where its decreasing value suggests that $\Delta_1$ could eventually close at even higher pressures.

Both the increase of $\Delta_2$ and $T_c$ is likely due to a pressure-induced increase in the charge carrier concentration at the interface that promotes interfacial superconductivity from an underdoped towards the optimally doped regime of its phase diagram. This is further confirmed by the decrease of the normal state resistance, which shows that the total charge carrier content increases throughout the heterostructure. In addition, it was reported that pressure has a strong influence on the topological surface states of $Bi_2Te_3$ [29]. Since the topological surface states are of crucial importance for the observation of interfacial superconductivity [10], this may further enhance the interfacial superconductivity and contribute to increasing the critical temperature. In the case of iron-based superconductors and the cuprates, as well as many other unconventional superconductors, the electronic density of states at the Fermi level can be altered by chemical substitution which drives superconductivity through a dome-shaped phase [30-32]. The application of high pressure has a very similar effect and increases the charge carrier concentration [33]. Above the superconducting transition temperature, the electrical transport of the heterostructure is largely dominated by the 140 nm thick layer of bulk $Fe_{1+y}Te$. The normal state resistivity at ambient pressure agrees well with literature data [10,14] for a similar large value of $y = 0.15 \pm 0.02$ as in our heterostructure, which shows a rather continuous insulator – metal transition in form of a resistivity maximum around 76 K at ambient pressure, instead of the sharper first-order resistivity jump for smaller $y$ values. The upturn of the resistance below ~20 K, which leads to the peak just above $T_c^{onset}$, was also observed for such samples and attributed to the magnetic scattering of the charge carriers on the interstitial Fe moments [25]. The paramagnetic high temperature insulating phase is almost unaffected by pressure, while the antiferromagnetically ordered low-temperature metallic phase becomes more metallic. Due to the latter behavior, it is difficult to estimate the pressure dependence of $T_{SDW}$. It appears to be almost unaffected by pressure, although it has been reported that in bulk $Fe_{1+y}Te$ the pressure decreases $T_{SDW}$ [27].

The pronounced peak above $T_c^{onset}$ is quickly suppressed by application of pressure. The scattering of charge carriers on interstitial Fe is therefore strongly suppressed and the more pronounced metallic behavior indicates a strong doping effect by pressure. We have previously shown that the interstitial Fe couples magnetically to the Fe moments in the $Fe_{1+y}Te$ layers [34]. The disappearance of the peak suggests that the magnetic moments of the interstitial Fe become more ordered, probably as a result of a stronger coupling to the antiferromagnetically ordered Fe moments within the $Fe_{1+y}Te$ layers. Another possibility is that the pressure induces a spatial ordering, or a kind of clustering of the interstitial Fe. This is illustrated by the dramatic sharpening of the superconducting transition under pressure. The interstitial Fe likely plays an important role in determining the normal conducting ground state of $Fe_{1+y}Te$. It provides strong pair breaking magnetic moments, and it also acts as a dopant [35]. Apart from this, a pressure-induced ordering can suppress the finite-size effect, which causes the broadening of the 2D Berezinski-Kosterlitz-Thouless transition [10]. The pressure-induced ordering or clustering of the interstitial Fe could reduce the finite-size effect, resulting in the sharper transition. This is further supported by the fact that both the $T_c^{onset}$ and $T_o$ remain higher than their initial values after releasing the pressure at the end of the experiment.

$T_c^{onset}$ and $T_o$ are enhanced under pressure, suggesting that the interface layer is still underdoped at ambient pressure and has the potential for a much higher critical temperature. The pressure has a similar effect to doping and is known to increase the charge carrier concentration of Fe-based superconductors [32]. This is certainly the case for the 140 nm thick bulk FeTe layer in our heterostructure [27] and thus also the interface, which results in its higher critical temperature. Since the TI certainly plays a decisive role in the emergence of superconductivity [10], there is probably also charge transfer from the TI to the interface.

The smaller gap $\Delta_1$ was attributed to a proximity-induced gap in the $Bi_2Te_3$ layer. Its pressure development thus shows that the strength of the proximity effect goes beyond a maximum. This can result from a subtle balance between the increase of $\Delta_2$ and a decreasing coherence length that pushes the interfacial superconductivity even further into the 2D limit with reduced superconducting thickness. On the other hand, it may also indicate that the TI is leaking charges to the interface layer, and thus plays a leading role in the doping effect, which is responsible for the occurrence of interfacial superconductivity.

## 5. Conclusions

In summary, we measured the electrical resistivity and point contact Andreev reflection spectroscopy on the novel interfacial superconductor $Bi_2Te_3$/$Fe_{1+y}Te$ under the influence of hydrostatic pressure. The temperature $T_o$ at which the zero resistance is established is increased from 4.0 K to 12.3 K with pressure rising from ambient condition to 5.91 kbar. We attribute this to a doping effect that drives interfacial superconductivity from the underdoped regime of its phase diagram towards optimal doping. In addition, we observed a dramatic sharpening of the resistive superconducting transition as well as a suppression of magnetic scattering above, ascribed to the pressure-induced ordering or clustering of interstitial Fe in the $Fe_{1+y}Te$ layer. The pressure development of the two superconducting gaps, attributable to the interfacial superconductivity and the proximity-induced superconductivity in the TI, respectively, suggests that the charge transfer from the TI layer to the interface occurs and demonstrates the important role of the TI layer in the mechanism of interfacial superconductivity.

## Acknowledgements

R. L. acknowledges stimulating discussions with Junwei Liu and Xi Dai. We thank U. Lampe and J. Song for technical support. This work was supported by grants from the Research Grants Council of the Hong Kong Special Administrative Region, China (16304515, SBI15SC10). Research at Washington University is supported by the National Science Foundation (NSF) through Grants No. DMR-1104742 and 1505345.

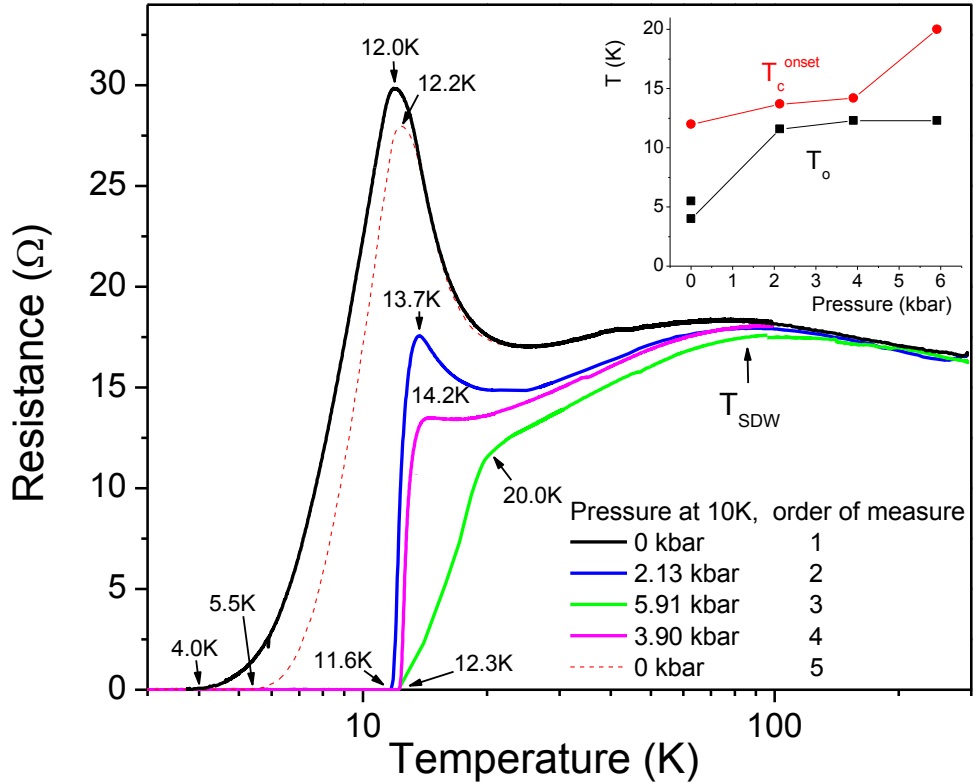

**Fig. 1**: The temperature dependent resistance measurement under pressures up to 5.91 Kbar. The temperature $T_o$ where the resistance reaches zero is enhanced from 4.0 K to 12.3 K by applying pressure, while the peak located at $T_c^{onset}$ is suppressed thus shifting the onset of the critical temperature up to 20 K at the highest pressure. $T_{SDW}$ marks approximately the transition below which the antiferromagnetic spin density wave is formed. The inset shows the pressure evolution of the critical temperatures $T_o$ (where zero resistance is reached) and $T_c^{onset}$.

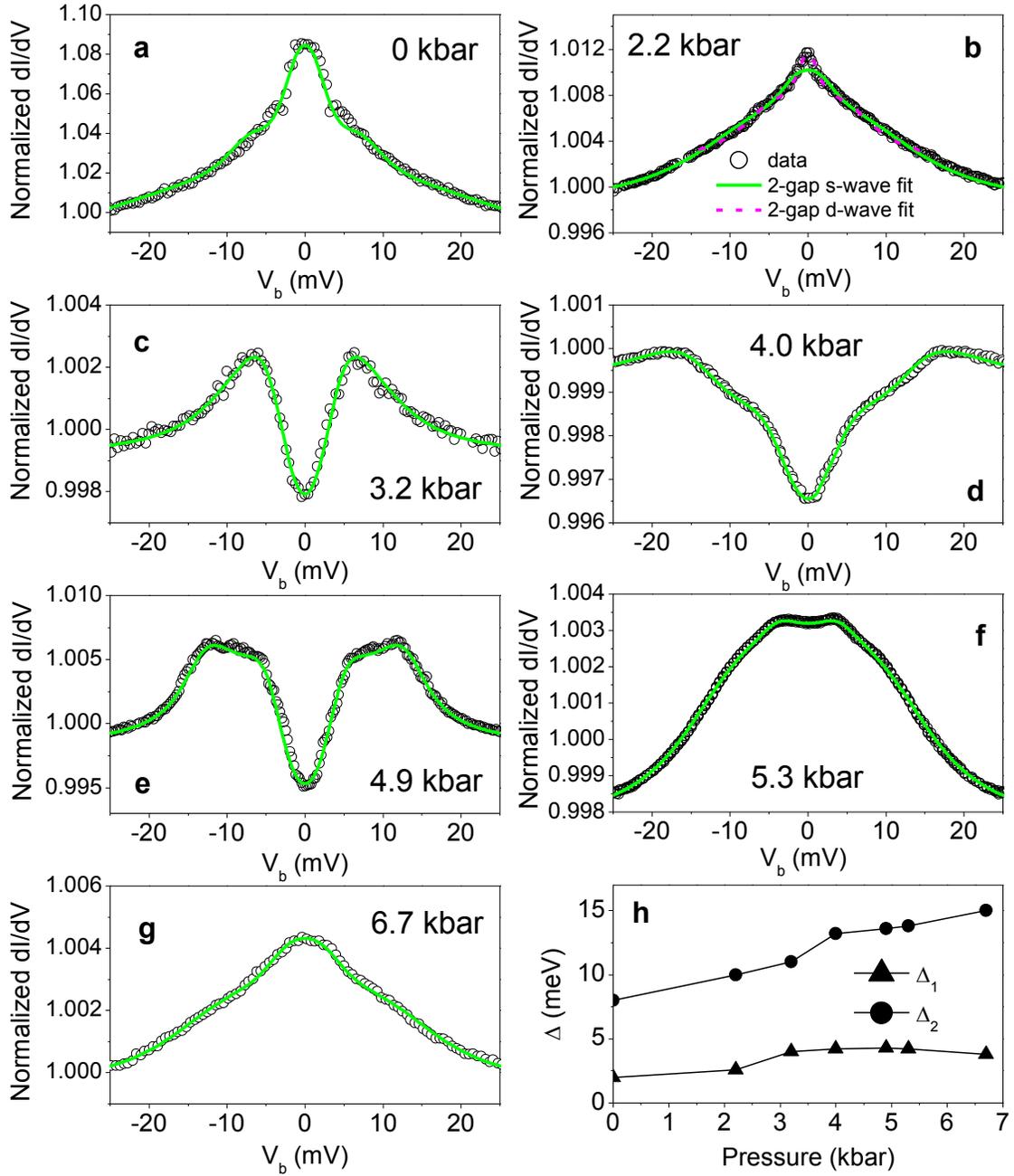

**Fig. 2**: **a-g**: Normalized point contact spectroscopy data (open circles) measured at 4.2 K at different pressures with a comparison of a two-gap s-wave BTK model fit (solid line). The dotted line in (**b**) is an additional two-gap d-wave BTK model fit along the nodal direction with the same gap parameters. **h**: Pressure dependence of the two superconducting gap values $\Delta_1$ and $\Delta_2$ extracted from the fits.